\documentclass[aps,prb,twocolumn,superscriptaddress,groupedaddress,floatfix,longbibliography]{revtex4-1}
\usepackage[utf8]{inputenc}

\usepackage{graphicx}  
\usepackage{dcolumn}   
\usepackage{bm}        
\usepackage{amssymb}   
\usepackage{amsmath}   
\usepackage{amsthm}    
\usepackage{xcolor}    
\usepackage{tikz}      
\usepackage{ifthen}    
\usepackage{subfigure}
\usepackage{xspace}
\usepackage[toc,page]{appendix}
\newcommand{\ket}[1]{{\vert {#1} \rangle}}

\newcommand{\swap}{{\text{\textbf{swap}}}}
\newcommand{\B}{{\mathcal{B}^{m}_{\mathbf{X}}}}

\newcounter{para}

\begin{document}

\title{
Entanglement Clustering  for ground-stateable quantum many-body states
}
       
\author{Michael Matty$^1$}
\email{mfm94@cornell.edu}
\author{Yi Zhang$^{1,2}$}
\author{T. Senthil$^3$}
\author{Eun-Ah Kim$^1$}
\email{eun-ah.kim@cornell.edu}

\affiliation{$^1$Department of Physics, Cornell University, Ithaca, New York 14853, USA}
\affiliation{$^2$International Center for Quantum Materials, Peking University, Beijing, 100871, China}
\affiliation{$^3$Department of Physics, Massachusetts Institute of Technology, Cambridge, MA, 02139}

\date{\today}

\begin{abstract}
Despite their fundamental importance in dictating the quantum mechanical properties of a system, ground states of many-body local quantum Hamiltonians form a set of measure zero in the many-body Hilbert space. Hence determining whether a given many-body quantum state is ground-stateable is a challenging task. 
Here we propose an unsupervised machine learning approach, dubbed the Entanglement Clustering (``EntanCl''),  to separate out ground-stateable wavefunctions from those that must be excited state wave functions using entanglement structure information. EntanCl uses snapshots of an ensemble of swap operators as input and projects this high dimensional data to two-dimensions, preserving important topological features of the data associated with distinct entanglement structure using the uniform manifold approximation and projection (UMAP). The projected data is then clustered using K-means clustering with $k=2$.
By applying EntanCl to two examples, a one-dimensional band insulator and the two-dimensional toric code, we demonstrate that EntanCl can successfully separate ground states from excited states with high computational efficiency. Being independent of a Hamiltonian and associated energy estimates, EntanCl offers a new paradigm for addressing quantum many-body wave functions in a computationally efficient manner.
\end{abstract}

\maketitle

\newpage

\section{Introduction}  
    Quantum many-body wave functions are complex objects, which encode a great deal
    of information. However, interpreting this information is difficult due to the 
    exponential number of parameters in the wave function and the need for a technique
    to interpret those parameters. In particular, we are interested in separating out wave functions that can be ground states
    of local Hamiltonians from the exponentially large space of all wave functions. Unfortunately, such ``ground-statable'' wavefunctions likely form a set of measure zero in the full many-body Hilbert space
        \cite{eisert:rmp2010a,page:prl1993a,foong:prl1994a,sen:prl1996a}. 
    Although the typical approach to wave functions is to measure their energies against a particular Hamiltonian of interest, such ranking by energy is subject to change when details of the Hamiltonian change. 
       
    As an alternative to resorting to a Hamiltonian, one could turn to entanglement properties. 
    In particular, given a
    partitioning of a system into two subregions $A$ and $B$, the scaling of the (Von Neumann) entanglement entropy  $S_A = -\text{Tr} \rho_A \ln \rho_A $ where $\rho_A$ is the 
    reduced density matrix of subregion $A$ can help determine groundstateability \cite{srednicki:prl1993a}. 
    Groundstateable wave functions typically exhibit $S_A$ that scales as the codimension 1 
    boundary of the cut between subregions $A$ and $B$ (area law), while that of
    non-groundstateable wave functions typically scales as a codimension 0 boundary (volume law).
    Such a distinction has indeed previously been used to distinguish groundstateable and
    non-groundstateable wave functions (see for example 
    \cite{vidmar:prl2018a,vidmar:prl2017a,keating:cmp2015a,storms:pr2014a,ares:jpmt2014a,alba:jsmte2009a,miao:apa2019a}).
    However, at a practical level, an investigation of the entanglement entropy scaling is often prohibitively expensive and the finite-size effects can make it challenging to declare area or volume law with confidence.  
    Clearly, a computationally efficient approach to separate out ground-stateable wave functions in an unbiased fashion is much desired. 

    Here we introduce ``EntanCl'' (Entanglement Custering), a machine learning approach designed to learn the entanglement structure of many-body quantum states and separate out ground-stateable states from rest of the Hilbert space in a computationally efficient yet unbiased manner. 
    Increasingly, the quantum condensed matter community is succssfully applying machine learning approaches to various 
    tasks such as phase recognition
    ~\cite{broecker:sr2017a,broecker:ap2017a,zhang:prl2017a,zhang:pr2017a,wang:pr2016a,carleo:s2017a,carrasquilla:np2017a,nieuwenburg:np2017a,beach:pr2018a,chng:pr2017a,chng:pr2018a,deng:pr2017b,liu:prl2018a,nieuwenburg:pr2018a,ohtsuki:jpsj2016a,schindler:pr2017a,wetzel:pr2017b,wetzel:pr2017a,yoshioka:pr2018a,venderley:prl2018a,matty:apa2019a},
    hypothesis tests on experimental data 
    ~\cite{ghosh:apa2019a,zhang:apa2018a},
    and compact representation of  many-body wave functions 
    ~\cite{cai:pr2018a,carleo:s2017a,chen:pr2018a,deng:pr2017a,deng:pr2017b,gao:nc2017a,huang:ap2017a,liu:pr2017b,nomura:pr2017a,schmitt:sp2018a,torlai:np2018a}. 
    A common feature among these 
    different problems that motivates the use machine learning approaches is the need to
    find structure in voluminous and complex data. However,
    the vast majority of the applications so far use supervised learning, which requires labeled training data and researchers' bias gets built into the labeling of the training data. 
    Without the pre-conceived notion of what makes a wave function groundstateable, we would like to separate out ground-stateable wavefunctions by learning the entanglement structure inherent in the many-body wave functions. For this, EntanCl uses Monte Carlo snapshots of the swap operator as the subsystem partition scans over the system. Then it employs 
    uniform manifold approximation and projection (UMAP)
    \cite{mcinnes:ap2018a} which is
    an \emph{unsupervised} ML approach of manifold learning in high-dimensional spaces to project the data down to a two-dimensional space. The final step of EntanCl is to cluster using K-means clustering.
    
    We will demonstrate the effectiveness of EntanCl by applying the method to many-body states associated with two specific models:
    a one-dimensional band insulator and Kitaev's 
    toric code \cite{kitaev:ap2006a} in two dimensions.
    The models are chosen to be representative of cases where the ground states
    and excited states are distinguished by entanglement structure, and are useful
    benchmarking cases because we know precisely what the ground states are.
    For any ML approach to data to be successful, it is critical to select relevant features to be fed into the ML algorithm.
    Motivated by the previously established importance of 
    entanglement properties in determining groundstateability, we will use an ensemble 
    of swap operators \cite{hastings:prl2010a} as feature selectors for our wave functions.

    The rest of the paper is organized as follows. 
    In section II, we introduce and describe the three steps of EntanCl. 
    In section III, we apply EntanCl to a simple, one-dimensional band insulator
    model and study the accuracy of our method in classifying wave functions.
    In section IV, we apply EntanCl to a strongly correlated problem: Kitaev's toric code\cite{kitaev:ap2006a}.
    In section V, we summarize our conclusions and discuss possible future applications.
   
\section{Methods}
\begin{figure}
        \centering
        \subfigure[]{
            \centering
            \includegraphics[width=\linewidth]{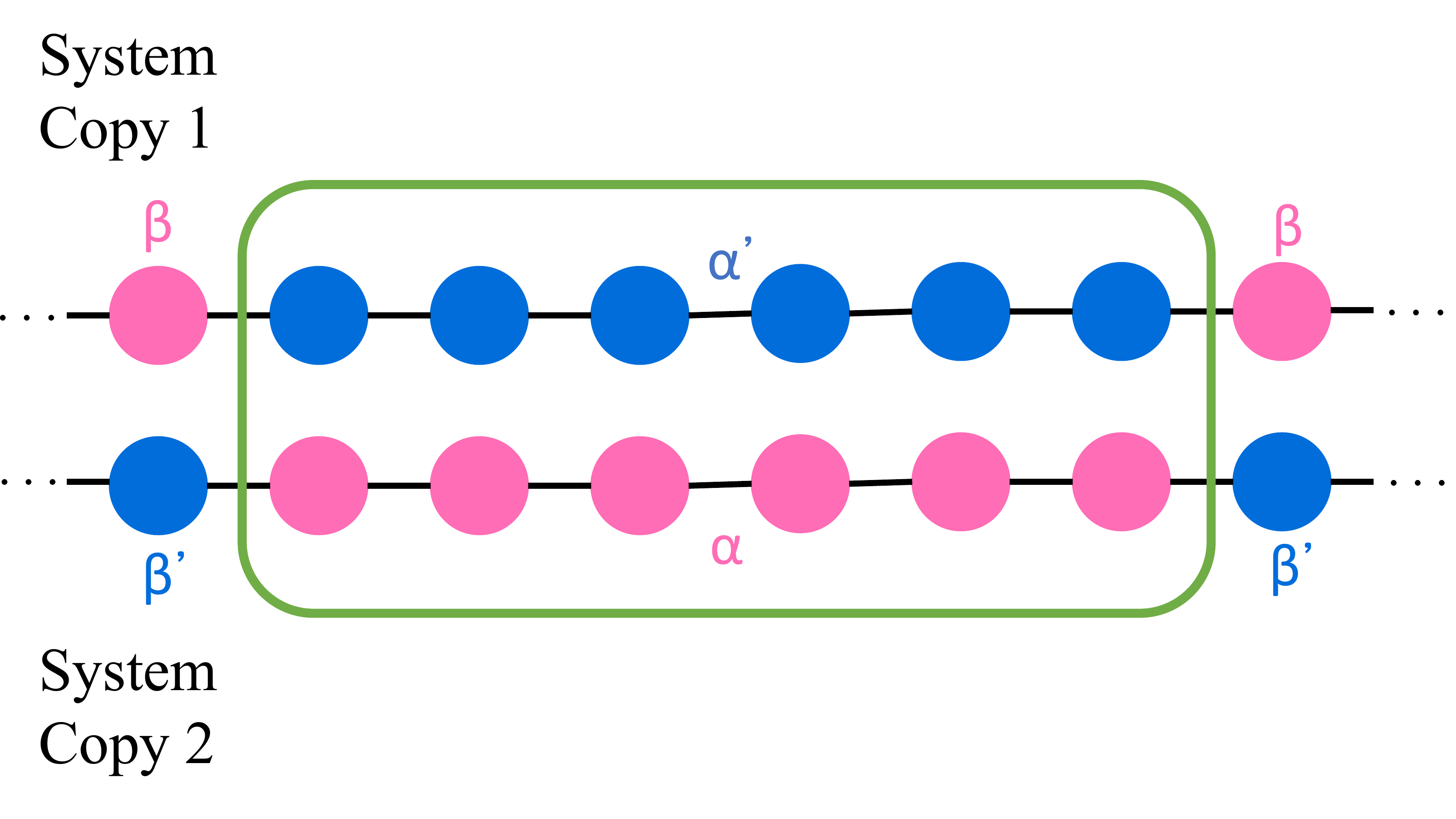}
        }
        \subfigure[]{
            \centering
            \includegraphics[width=\linewidth]{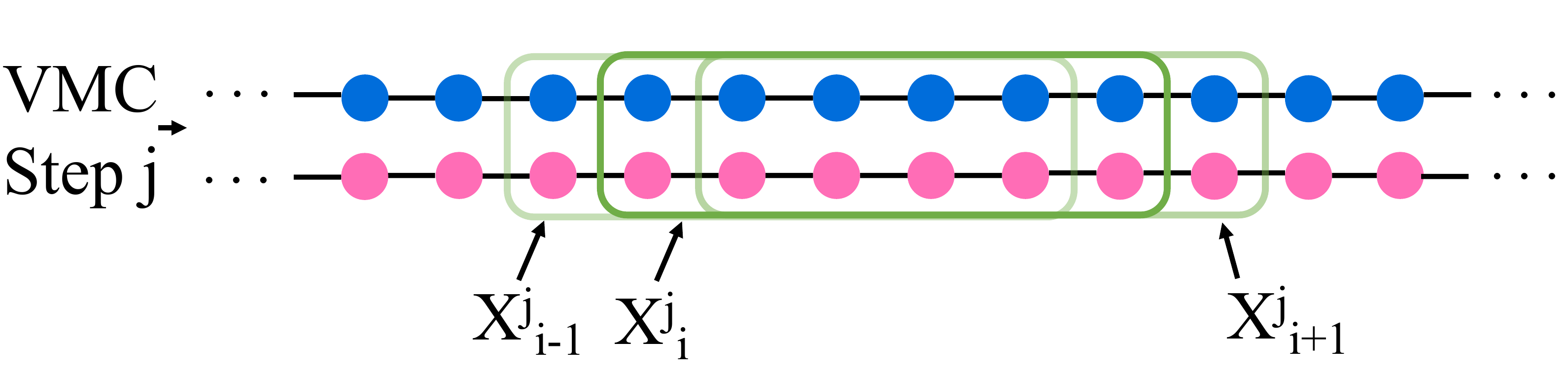}
        }
        \caption{
        (a) Schematic depiction of the action of the swap operator on a subsystem
        \textbf{A}. The quantum numbers $\alpha$ describe the subsystem and 
        $\beta$ describe the remainder of the system. Since $\swap$ acts on a 
        doubled Hilbert space, we denote the quantum numbers belonging to one
        copy by primed variables and to the other by unprimed variables.
        The operator $\swap_A$ switches the primed and unprimed variables within
        the region \textbf{A}.
        (b) Illustration of our data collection procedure. At each VMC step $j$,
        we collect swap data from a collection of
        subsystems $A_i$ and store each in a vector $\vec{X}^j$ at index $i$. The collection
        of $\vec{X}^j$'s forms our complete dataset $\mathbf{X}$.}
        \label{fig:data_schematic}
    \end{figure}

    EntanCl consists of three steps. The first step is to construct the input data of swap operator snapshots. 
    In search of the right feature selection approach, we are inspired by the use of the 
    swap operator in calculating Renyi entropies \cite{hastings:prl2010a}. 
    The action of the swap operator is illustrated in fig.~\ref{fig:data_schematic}.
    The expectation
    value of the swap operator in the state $\ket{\Psi}=\sum_{\alpha,\beta}C_{\alpha\beta}
    \ket{\alpha\beta}$ is given by
        \begin{align}
        \label{eqn:matelt}
        \langle \swap_A \rangle = e^{-S_2} = 
        \sum\limits_{\alpha,\beta,\alpha',\beta'}
        \vert C_{\alpha\beta} \vert^2 \vert 
        C_{\alpha' \beta'} \vert^2
        \frac{C_{\alpha'\beta}C_{\alpha\beta'}}
        {C_{\alpha\beta}C_{\alpha'\beta'}}
    \end{align}
    where $S_2$ denotes the second Renyi entropy, $A$ denotes a subsystem, the quantum numbers 
    $\alpha$ describe subsystem $A$, and $\beta$ describe the remainder of the system. 
    We will \emph{not} take the expectation value, however. 
    Instead, we will variationally sample the swap data for
    $\ket{\Psi}=\sum_{\alpha,\beta}C_{\alpha\beta}
    \ket{\alpha\beta}$ according to eq.~(\ref{eqn:matelt}),
    where $\vert C_{\alpha\beta}\vert^2 \times \vert
    C_{\alpha'\beta'}\vert^2$ plays the role of the sampling weights.
    In order to acquire more comprehensive data across the system,
    we will consider many subsystems $A_i$ to form an ensemble of swap operators
    $\{\swap_{A_i}\}$. 
    
    As we sample the swap data  
    with variational Monte Carlo (VMC), we build up a 
    collection of vectors 
    $\mathbf{X} = \{\vec{X}^j\}$
    (c.f. fig.~\ref{fig:data_schematic})
    where at index $i$, $\vec{X}^j$
    contains the data ${C_{\alpha'\beta}C_{\alpha\beta'}} /
    {C_{\alpha\beta}C_{\alpha'\beta'}}$
    sampled from $\swap_{A_i}$ at VMC step $j$. 
    The dimensionality of our data is precisely the
    number of subsystems $A_i$ we choose to consider. This will be order hundreds of dimensions
    for the band insulator and thousands for the toric code.
    We thus have a high-dimensional data set $\mathbf{X}$ that contains entanglement
    information about the wave function $\ket{\Psi}$.

    The second step of EntanCl is to project the input data living in the high dimensional space (typically 
    hundreds or thousands of dimensions) down to two-dimensional space in which clustering can be visualized.
    Typical applications of unsupervised ML to high-dimensional data sets involve visualizing the data in a low-dimensional space via dimensional reduction.
    Dimensional reduction algorithms (such as those described in 
    refs.~\cite{tang:2016a,maaten:jmlr2008a,coifman:acha2006a,belkin:2002a,tenenbaum:s2000a,sammon:itc1969a,kruskal:p1964a,hotelling:jep1933a})
    vary in the way that they approximate the high-dimensional
    manifold populated by the data and what features of that manifold 
    they try to preserve
    under projection to the low-dimensional space. We are interested in an algorithm
    that  will allow us to visualize the cluster structure in our swap data set $\mathbf{X}$. 
    This is because we expect that those $\vec{X}^j$ obtained
    from groundstateable and non-groundstateable wave functions
    will appear as two separate clusters due to differing entanglement structure. 
    
\begin{figure}
        \centering
        \includegraphics[width=\linewidth]{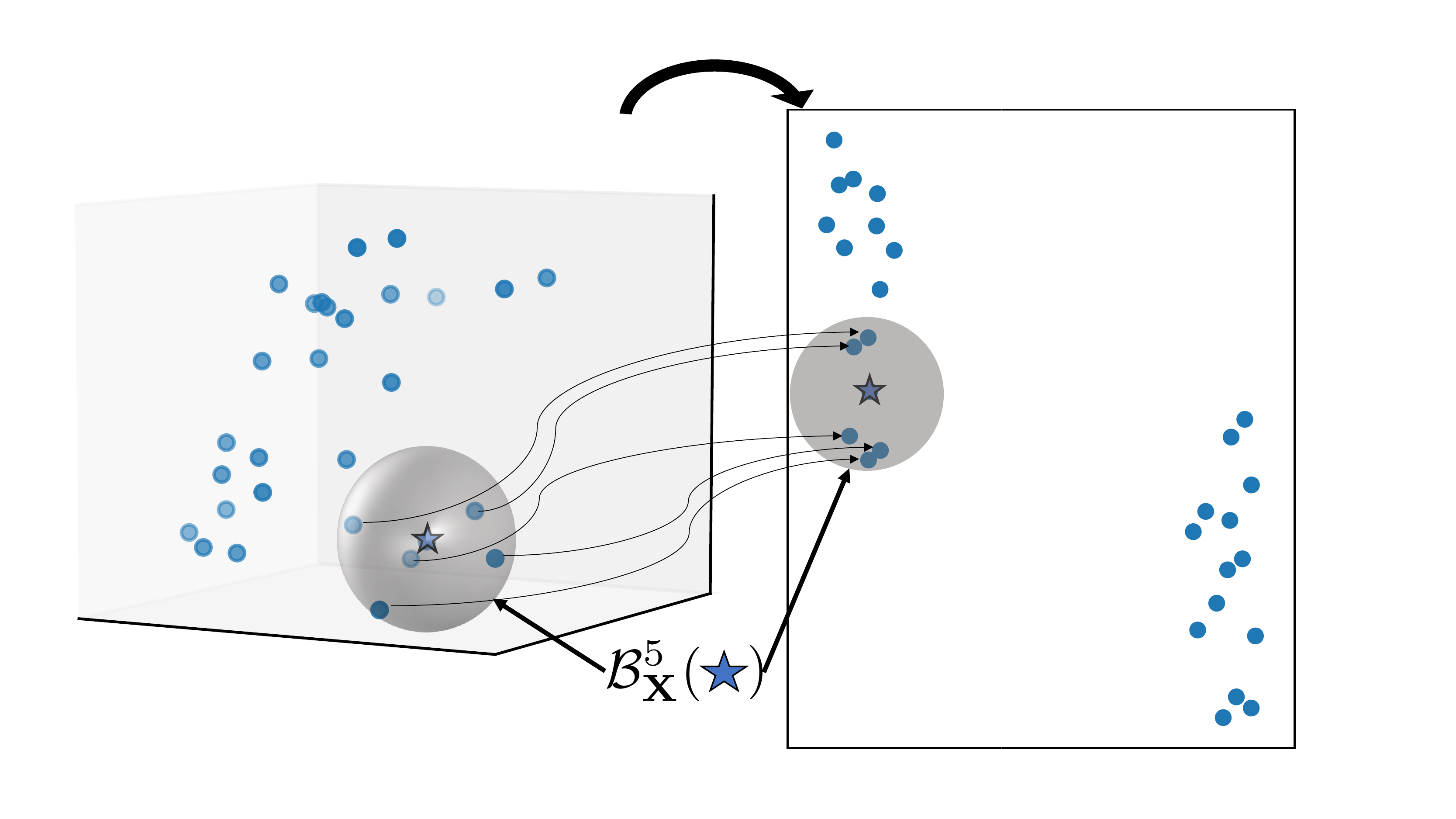}
        \caption{
        Schematic illustration of "neighborhood structure" preservation,
        projecting points in three dimensions to two.
        The five nearest neighbors of the star are found by application of
        $\mathcal{B}_\mathbf{X}^5$. After projection, we can see that 
        the five nearest neighbors of the point marked with a star remain its 
        five nearest neighbors. Moreover, by preserving local neighborhoods, 
        we have discovered two distinct clusters in the high dimensional data.
        For this example, the projection was done by UMAP.}
        \label{fig:umap_schematic}
    \end{figure}

    We can view clusters from a neighborhood perspective. 
    As an example, in fig.~\ref{fig:umap_schematic} we consider three dimensional data 
    consisting of two clusters:
    15 points randomly generated on the upper hemisphere of a unit
    radius sphere and 15 generated on the lower hemisphere. Gaussian noise
    is applied to the coordinates of the points. We then project the points down to 
    two dimensions so as to preserve their local neighborhood structure. In this case
    we use UMAP to do the projection.
    On the right hand panel of fig.~\ref{fig:umap_schematic}, we can see that in each
    of the two clusters, the local neighborhoods of each point are entirely contained within
    the same cluster as the point. To emphasize this, we illustrate a local neighborhood of size five
    around the point marked by a star.
    From this we can infer that preserving
    local neighborhood structure also preserves cluster structure. Formally, 
    define a function $\B$ such that $\B(\vec{X}_*) \subseteq \mathbf{X}$ is the set of 
    the $m$ nearest neighbors of $\vec{X}_*$ in $\mathbf{X}$. A cluster is then
    a subset $\mathbf{C} \subseteq \mathbf{X}$
    such that $\B(\vec{C} \in \mathbf{C}) \subseteq \mathbf{C}$. 
    For visualizing clusters, a natural choice for a dimensional reduction algorithm is then
    one that preserves neighborhoods after projection. 
    
    Algorithms that preserve neighborhood structure
        \cite{tang:2016a,maaten:jmlr2008a,coifman:acha2006a,belkin:2002a,tenenbaum:s2000a}
        try to find a
        mapping $\mathcal{P}$ from the $D$-dimensional data space to $\mathbb{R}^d$ 
        (again, $\mathbb{R}^2$ for us), such that 
        $\mathcal{P} \circ \B = \mathcal{B}^m_{\mathbf{\mathcal{P}(X)}}\circ\mathcal{P}$
        where $\circ$ denotes the usual composition of mappings. 
        Observe that preserving neighborhoods entails not
        only keeping points within a cluster nearby, but keeping points in separate clusters far away
        from each other. 
        Common algorithms accomplish this by taking as input a hyperparameter that defines an
        estimated neighborhood or cluster size, related to the $m$ in our definition of $\B$. 
        These algorithms treat the effective distance between
        points outside of a neighborhood as extremely (or sometimes infinitely) far away.
        One must be sure to choose this hyperparameter large enough (based on the density of
        the data) that spurious clusters do not appear in the projected data. That is to say
        that the intersection of the neighborhoods $\B$ need to contain the entire, true cluster.
        For our purposes, we use UMAP, which has previously found use in 
        biology \cite{becht:b2018a,diaz-papkovich:b2019a,park:b2018a,oetjen:i2018a,bagger:nar2018a,clark:b2018a,kulkarni:cob2019a,la-manno:n2018a,wolf:2018a}, 
        materials engineering \cite{fuhrimann:2018a}, 
        and machine learning \cite{blomqvist:apa2018a,gaujac:apa2018a,escolano:apa2018a}, but has had
        limited use in quantum matter \cite{li:cm2019a}.
        For more details about how UMAP in particular works, see appendix A.
        We choose UMAP from the various unsupervised ML algorithms that seek to preserve neighborhood structures for two reasons.
        Firstly, it led to the most clear projected clustering for our purposes. Secondly, in contrast
        to other algorithms like tSNE, UMAP provides us with a transferable mapping
        that can be applied immediately to new data without rerunning UMAP.

    The final step of EntanCl is to intrepret the learned UMAP output using $k$-means
    clustering. 
    $K$-means clustering 
    partitions a set of data points into $k$ clusters by placing $k$ cluster 
    means (centroids) in a way that minimizes the sum of squared distances from each
    data point to its nearest centroid.  A $(k=2)$-means clustering thus naturally allows 
    us to classify (non-)groundstateable wave functions in the 2-D projected space.
    For our test cases where we know which cluster corresponds to each type of 
    wave function, we define a metric of accuracy given by assignment to the correct centroid.

\section{Band Insulator}
    To establish EntanCl on a simple, known model,
    we first study a one-dimensional band insulator.
    This model is described by the Hamiltonian
    \begin{align}
        \mathcal{H} = \sum\limits_i (t_1 b_i^\dagger a_i +
        t_2 a^\dagger_{i+1} b_i) + \text{h.c.}.
    \end{align}
    This model has two bands with energy gap 
    $\Delta E \sim \vert t_2 - t_1 \vert$, and we consider
    the case of half filling. 
    We report results in terms of the dimensionless, normalized gap $t \equiv \vert t_2 - t_1 \vert / t_1$.
    The ground state Slater determinant wave function of the
    half filled system corresponds to completely filling the
    lower band. The non-groundstateable eigenstates
    we consider have some
    fixed density $n_{ex} \equiv N_{ex}/L$ of randomly chosen $k$-points
    promoted to the upper band, where $L$ is the system size.
    This model gives us a testbed to identify ground state wave functions
    and non-groundstateable wave functions in the parameter space of
    energy gap $\Delta E$ and excited $k$-point density $n_{ex}$.
    
    The ensemble of swap operators we use in this case is the set of all contiguous
    length six subsystems of an $L = 100$ chain.
    Our data set $\mathbf{X}$ consists of 1000, 100-dimensional swap vectors
    $\vec{X}^j$ corresponding to the ground state and 1500 corresponding to a non-groundstateable
    wave function. 
    We choose an uneven ratio of swap data from the two classes to illustrate
    that a symmetric amount of data is nonessential to our technique.
    We project the data to two dimensions via UMAP and assign the projected
    data points to clusters with $k$-means. Since we know which swap data points came from
   (non-)groundstateable wave functions, we also calculate the accuracy.
   
    \begin{figure*}
        \centering
        \subfigure[]{
            \centering
            \includegraphics[height=1.6in]{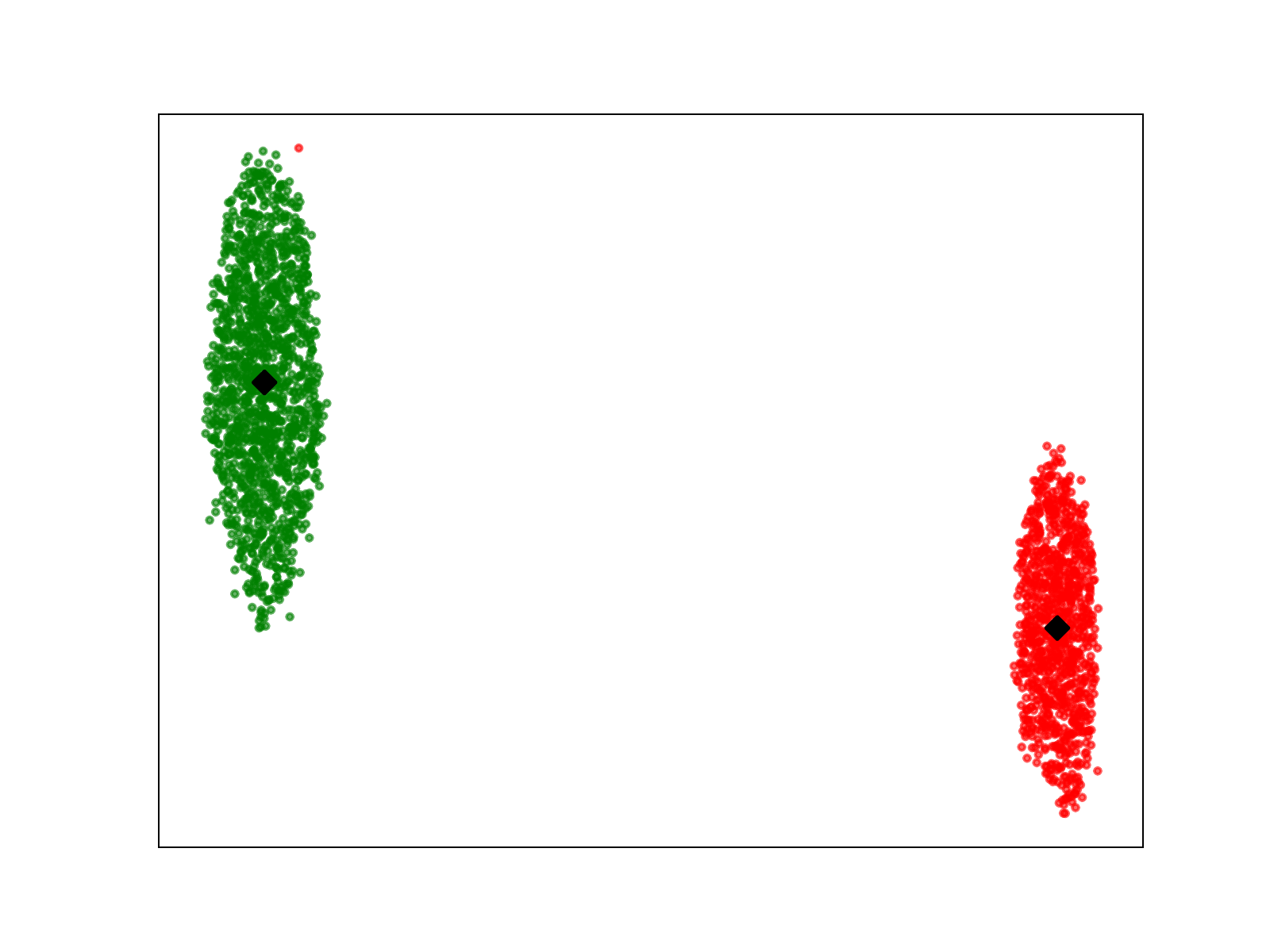}
        }
        \subfigure[]{
            \centering
            \includegraphics[height=1.7in]{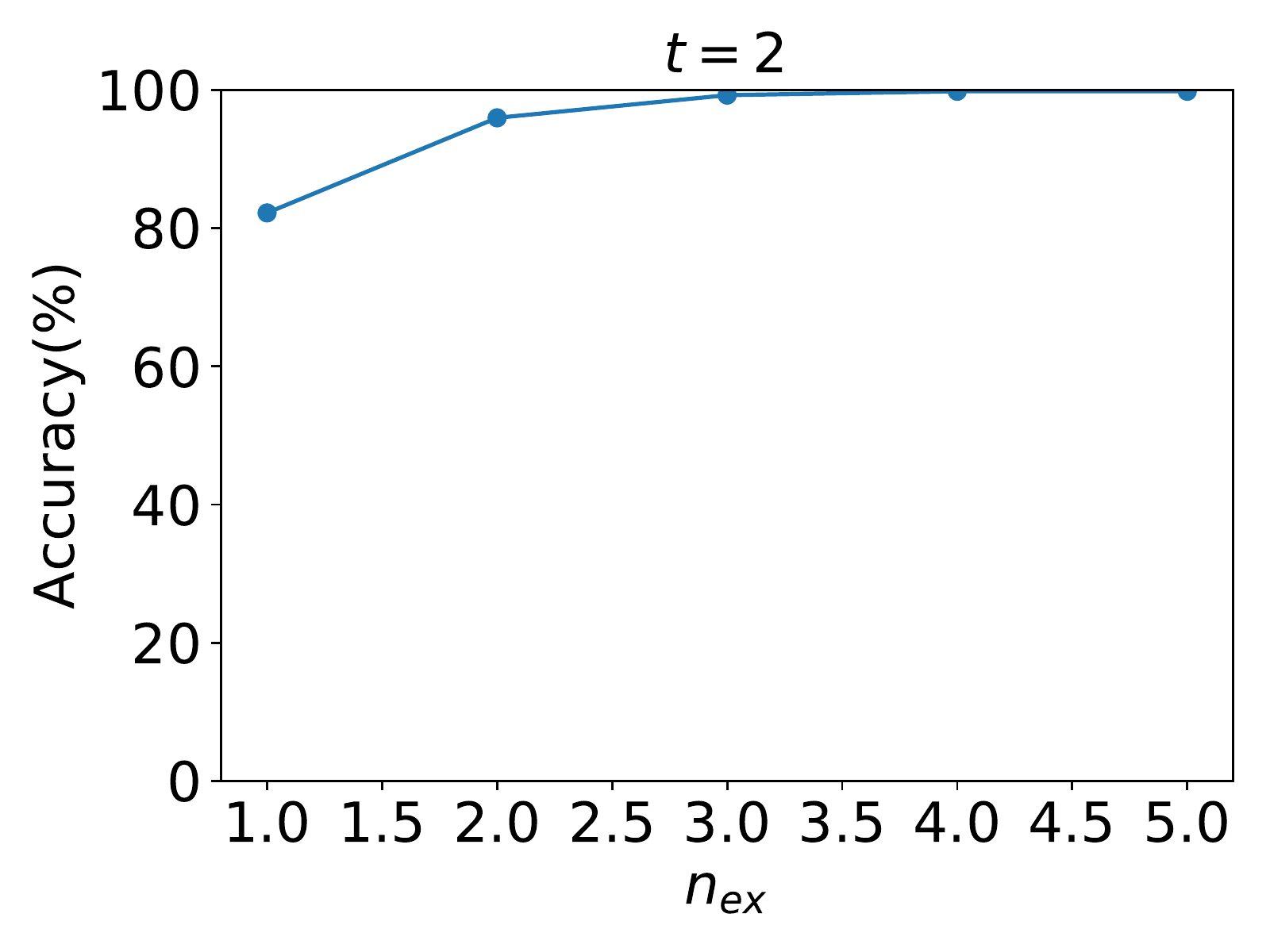}
        }
        \subfigure[]{
            \centering
            \includegraphics[height=1.7in]{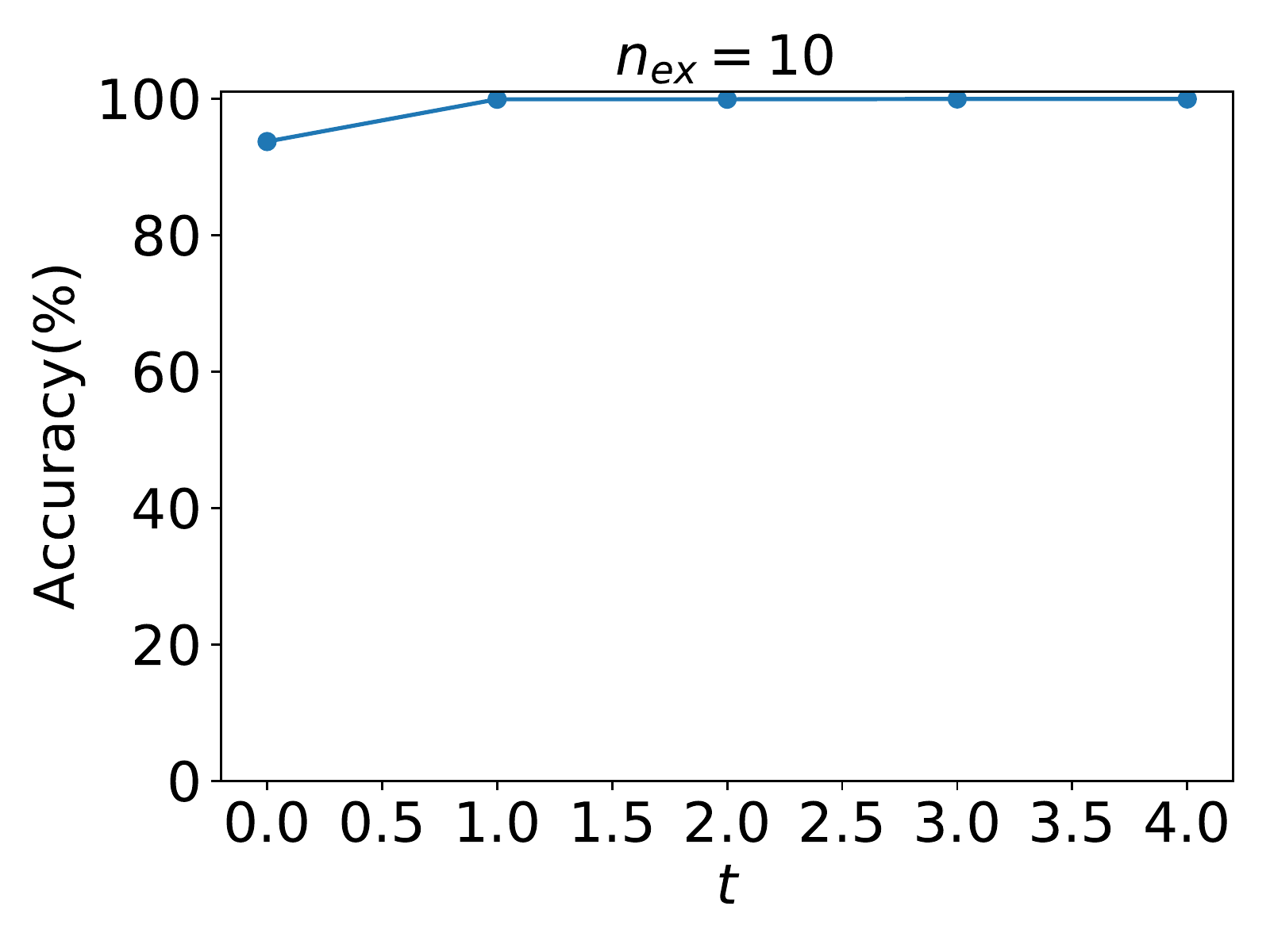}
        }
        \caption{
        (a) UMAP projection of swap data obtained from wave functions for
        band insulator model. Red dots correspond to swap data from a groundstateable
        wave function. Green dots correspond to swap data from a non-groundstateable 
        wave function with $n_{ex} = 3\%$ and $t = 2$. Black diamonds
        denote the $(k=2)$-means clustering centroids. This case has accuracy $96.52\%$.
        We also show the accuracy as a function of (b) excitation density $n_{ex}$ 
        at normalized energy gap $t = 2.0$ and (c) $t$ at $n_{ex} = 10\%$. 
        In both cases, accuracy increases as a function of
        the relevant parameter, and moreover, stays relatively high at the minimum 
        possible value.}
        \label{fig:band_insulator}
    \end{figure*}

    Our results are shown in figure \ref{fig:band_insulator}. 
    Fig.~\ref{fig:band_insulator} (a), corresponds to a projection with the normalized gap $t = 2$ and excitation density
    $n_{ex} = 3\%$. In this case one can clearly see the success of EntanCl: the
    data corresponding to the groundstateable wave function (red) and the non-groundstateable
    wave function (green) appear as two well separated clusters. This case corresponds to
    an accuracy of $99.12\%$. In fig.~\ref{fig:band_insulator}(b,c) we can see that as both $t$ and 
    excitation density increase, the accuracy also increases.
    This makes sense: as both $t$ and $n_{ex}$ increase, the excited state becomes more entangled
    compared to the ground state as the entanglement entropy scaling transitions from 
    area law to volume law.
    Moreover, the accuracy stays high
    even at the lowest possible $n_{ex}$ ($80.00\%$ for $t = 2$) and for a gapless system
    ($90.03\%$ for $n_{ex} = 10\%$).  This demonstrates that EntanCl is a viable method of identifying
    the differing entanglement structure in groundstateable and non-groundstateable.
    
\begin{figure}
        \centering
        \includegraphics[width=\linewidth]{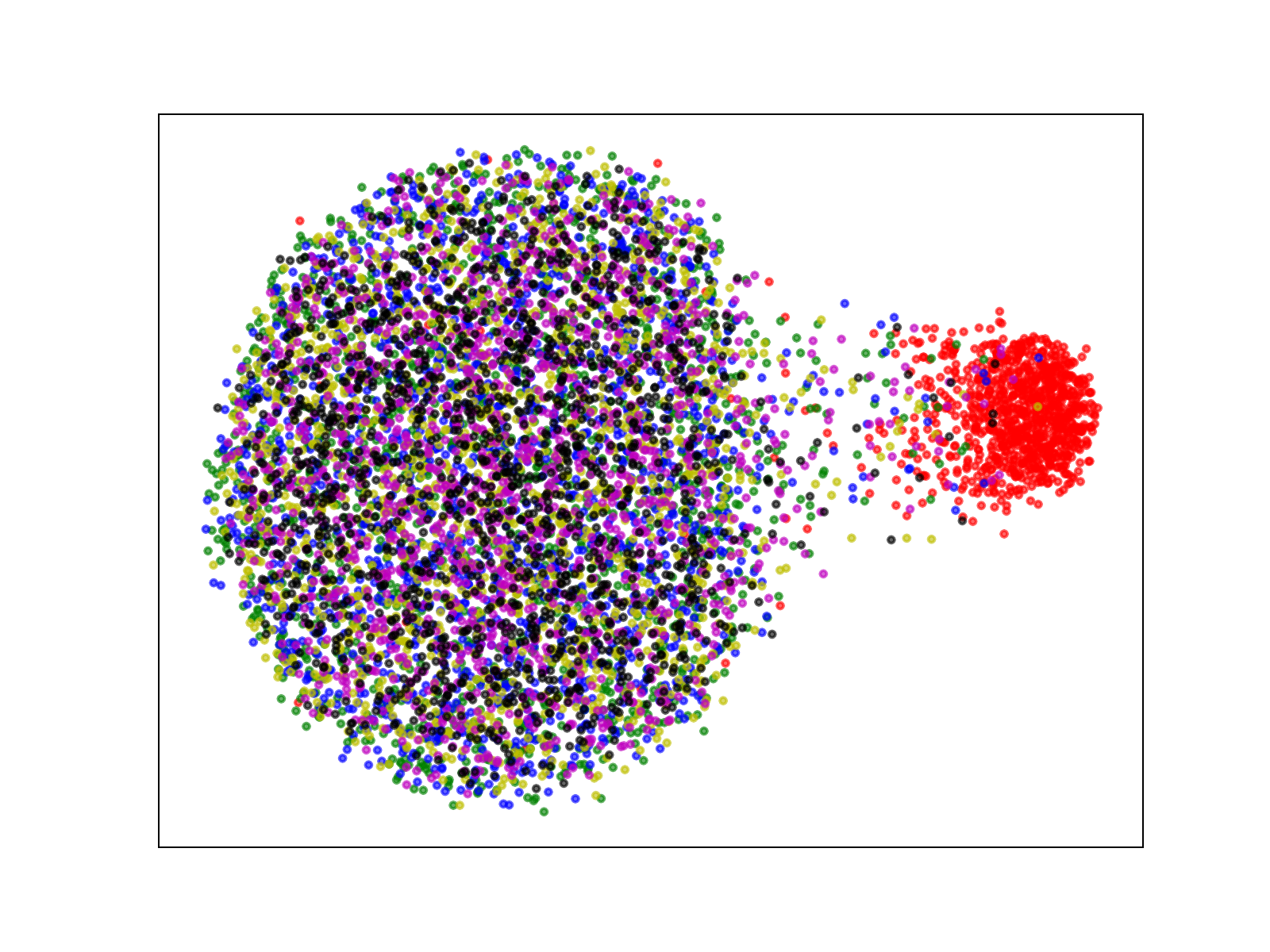}
        \caption{UMAP projection of swap data from band insulator wave functions 
        at gap $t = 2$ and excitation density $n_{ex} = 2\%$. The UMAP projection
        was trained using the ground state and a single excited state configuration
        (i.e. single choice of excited $k$-points). We then transfer the mapping
        to four more excited state configurations and display the results simultaneously.
        The ground state data are shown in red, the other colors correspond to 
        various excited state configurations. Clearly, subsequent excited states
        cluster together with each other, and more importantly all cluster separately from
        the ground state. 
        }
        \label{fig:bi_multicluster}
    \end{figure}

        The learned UMAP projection is transferrable. 
        In fig.~\ref{fig:bi_multicluster} we illustrate the results of transferring the UMAP
        projection trained on swap data obtained from the groundstateable wave function 
        and a single non-groundstateable wave function (i.e. single choice of excited $k$-points)
        with $t = 2$ and $n_{ex} = 2\%$ to 
        four more non-groundstateable wave functions with the same $t$ and $n_{ex}$. 
        We collect 1000 MC samples for the groundstateable wave function and 
        1500 for each non-groundstateable wave function.
        The projection map clusters all the data from non-groundstateable wave functions together, 
        away from the data from the groundstateable wave function. 
        The accuracy in this cas is $84.4\%$, lower than the $96.6\%$ in fig.~\ref{fig:band_insulator}(b) for two
        wave functions. This is because most of the error is non-groundstateable data being misclassified as
        groundstateable. Increasing the amount of data collected from the groundstateable wave function would 
        increase the accuracy.
        These results show that the structure that UMAP is learning generalizes well.
    
\section{Toric Code}

    \begin{figure*}
        \centering
        \subfigure[]{
            \includegraphics[width=0.3\linewidth]{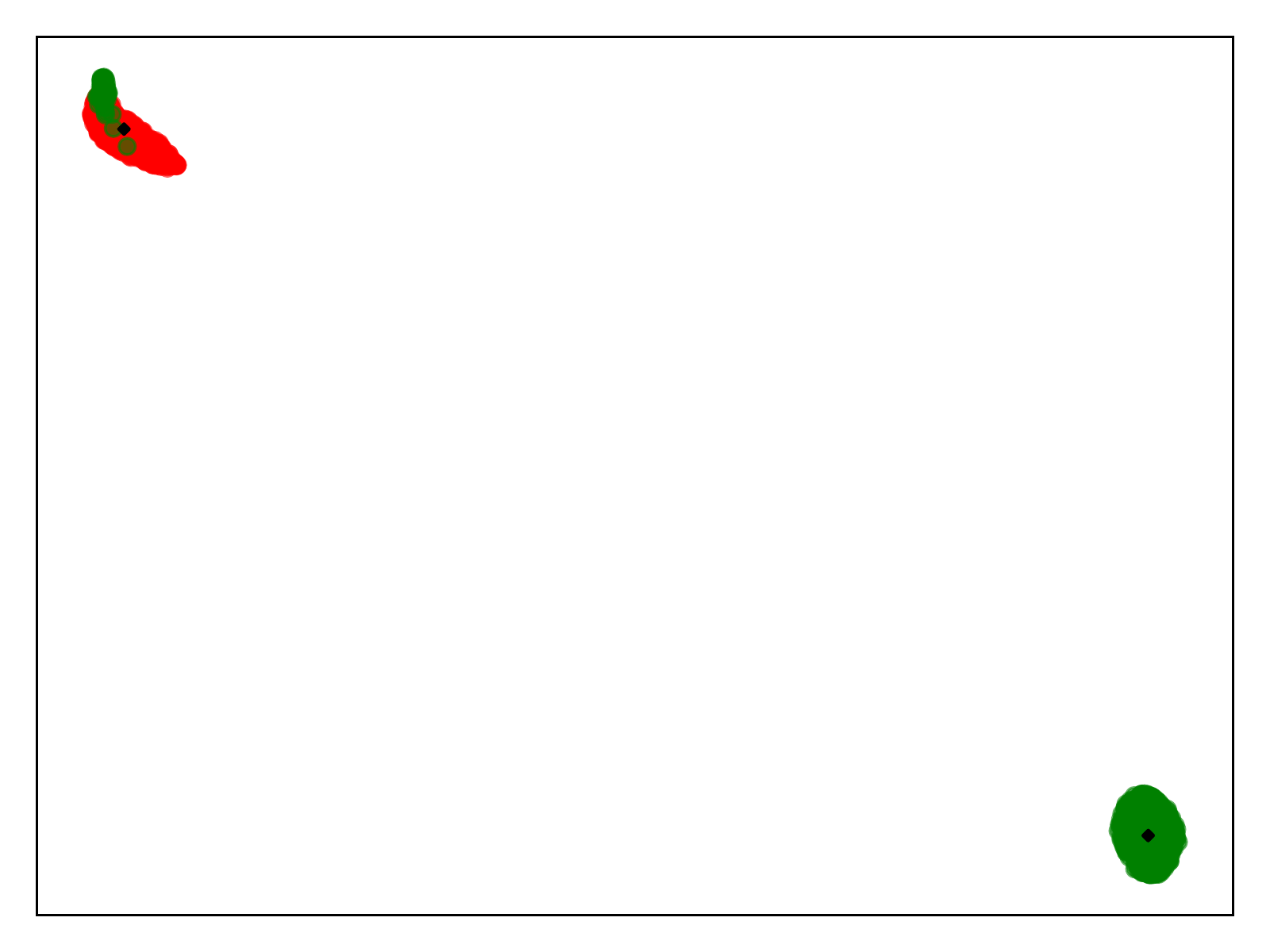}
        }
        \subfigure[]{
            \includegraphics[width=0.3\linewidth]{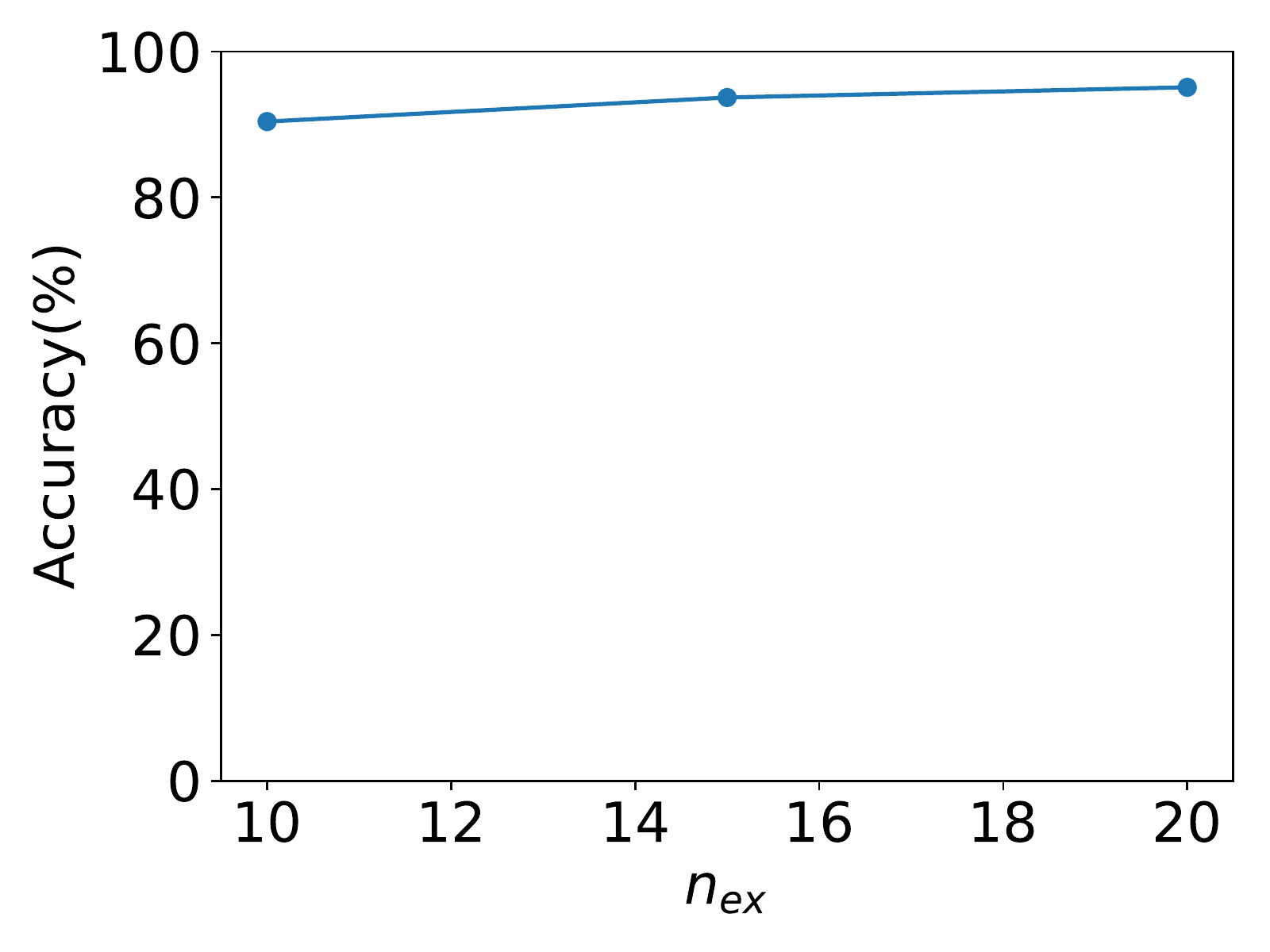}
        }
        \subfigure[]{
            \includegraphics[width=0.3\linewidth]{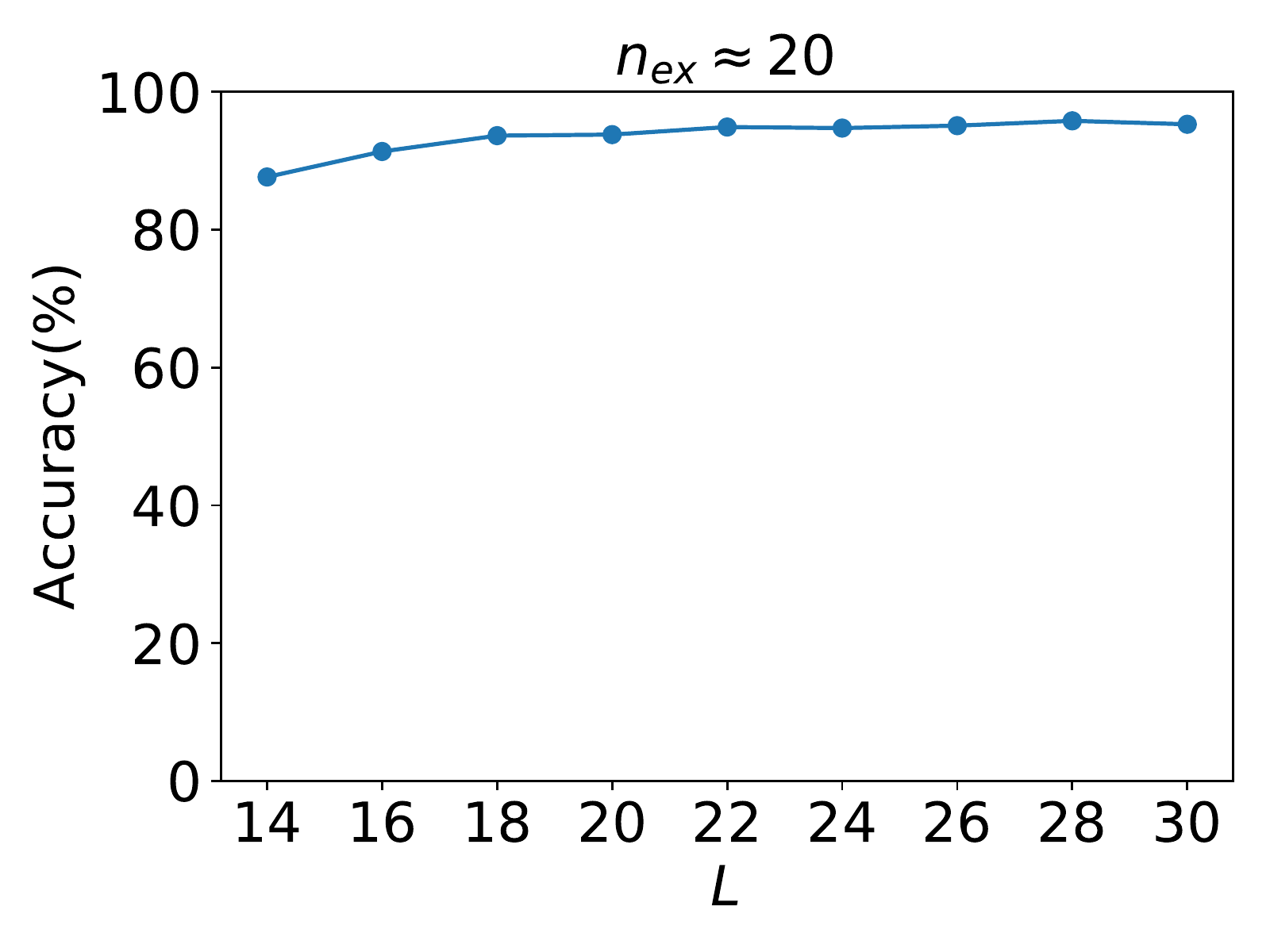}
        }
        \caption{(a) UMAP projection of swap data obtained from wave functions for
        the toric code. Red dots again correspond to swap data from groundstateable
        wave functions. Green dots correspond to swap data from a non-groundstateable
        wave function from a lattice with linear dimension $L=25$ with spinon density
        $n_{ex} = 20\%$. Black diamonds
        denote the $(k=2)$-means clustering centroids and this case correponds to accuracy
        $95.91\%$. (b) The accuracy at a fixed lattice size grows with $n_{ex}$, as expected.
        (c) Classification accuracy for UMAP projection of toric code wave functions
        as a function of lattice linear dimension.
        Data shown is at spinon density $\sim 20\%$. Accuracy increases with system size and plateaus around $95\%$. Slight non-monotonicity near the plateau 
        is expected because $n_{ex}$ must
        be an even integer and is therefore not exactly $20\%$ for all lattice sizes.}
        \label{fig:tc_results}
    \end{figure*}

    We now turn to a two-dimensional example: Kitaev's toric code \cite{kitaev:ap2006a}.
    This is a strongly interacting system whose ground state has topological order,
    and because it is exactly solvable, we will be able to assess the accuracy of EntanCl.
    This model is defined
    on a square lattice with spin-$1/2$ variables living on the edges.
    The wave functions that we will consider in this case are eigenstates of the Hamiltonian
    \begin{align}
        \mathcal{H} &= -\sum\limits_\square A_\square -\sum\limits_v B_v
    \end{align}
    where the operators 
    \begin{align}
        A_\square = \prod\limits_{i \in \square} \sigma^x_i,\ 
        B_v = \prod\limits_{i \in \partial v} \sigma^z_i
    \end{align}
    are defined as the product of pauli $\sigma^x$ operators around a plaquette and
    $\sigma^z$ operators on the edges incident on a vertex $v$ respectively.  Note that
    we will be working in the $\sigma^z$ basis.
    
    The ground state wave function we will consider is the equal amplitude superposition
    of all lattice configurations of closed loops in
    the trivial homology class.
    \footnote{A loop is a closed, connected path of edges with the
    same $\sigma_z$ eigenvalue, where at least one vertex that intersects the path has
    two edges of each $\sigma_z$ eigenvalue incident on it.}
    The non-groundstateable wave functions we will consider are equal amplitude superpositions
    of all states with a 
    fixed spinon density (also allowing closed
    loops) where a spinon is a vertex $v$ with $B_v = -1$. 
    Note that this does not correspond to fixed spinon \emph{locations}, as such
    wave functions could be made ground states by simply flipping the sign of the $B_v$'s 
    corresponding to the spinon locations.
    With this model, we will classify wave functions at different values of our
    control parameter: the spinon density $n_{ex}$.
    
    We collect swap data at 1000 uncorrelated VMC time steps for each wave function
    we consider. The ensemble of swap operators we use in this case consists
    of all rectangular subregions of the lattice, which grows with the
    linear dimension of the lattice $L$ as $L^4$.
    Due to the massively increased dimensionality of the swap data in this case,
    we add a preprocessing step to compress the data volume for RAM storage, 
    especially for larger system sizes.
    We average the swap data for a fixed subsystem width and height over all basepoints
    for the subsystem.  This reduces the dimensionality of the data to $L^2$, which is
    sufficiently tractable for our purposes.
    With this addition to our analysis, we can project the swap data to 
    two dimensions via UMAP.
    \footnote{
        Note that in this case all swap matrix elements are either 0 or 1, leading to
        occasionally redundant $\vec{X}_k$. We take only unique $\vec{X}_k$ here to avoid
        artificial clusters in the UMAP, but account for the multiplicity in error calculations.
    }
    
    Our results for the toric code are shown in fig.~\ref{fig:tc_results}.
    We find that we can achieve $95.91\%$ accuracy for $n_{ex} = 20\%$ for a lattice with
    linear dimension $L = 25$ as shown in fig.~\ref{fig:tc_results}(a).
    For a lattice with linear dimension $L = 35$ we get accuracy $99.1\%$ even at
    $n_{ex} = 5\%$.
    Once again, for
    this high accuracy case, the success of the clustering is remarkably clear. In 
    fig.~\ref{fig:tc_results}(b), we can see that the accuracy also increases
    with $n_{ex}$ as we would expect. Moreover, we do not need such a large system to
    achieve good accuracy. We can see in fig.~\ref{fig:tc_results}(c) that 
    for $n_{ex} = 20\%$, the accuracy of the projection
    is over $90\%$ already at $L = 16$.
    
    \begin{figure}
        \centering
        \includegraphics[width=\linewidth]{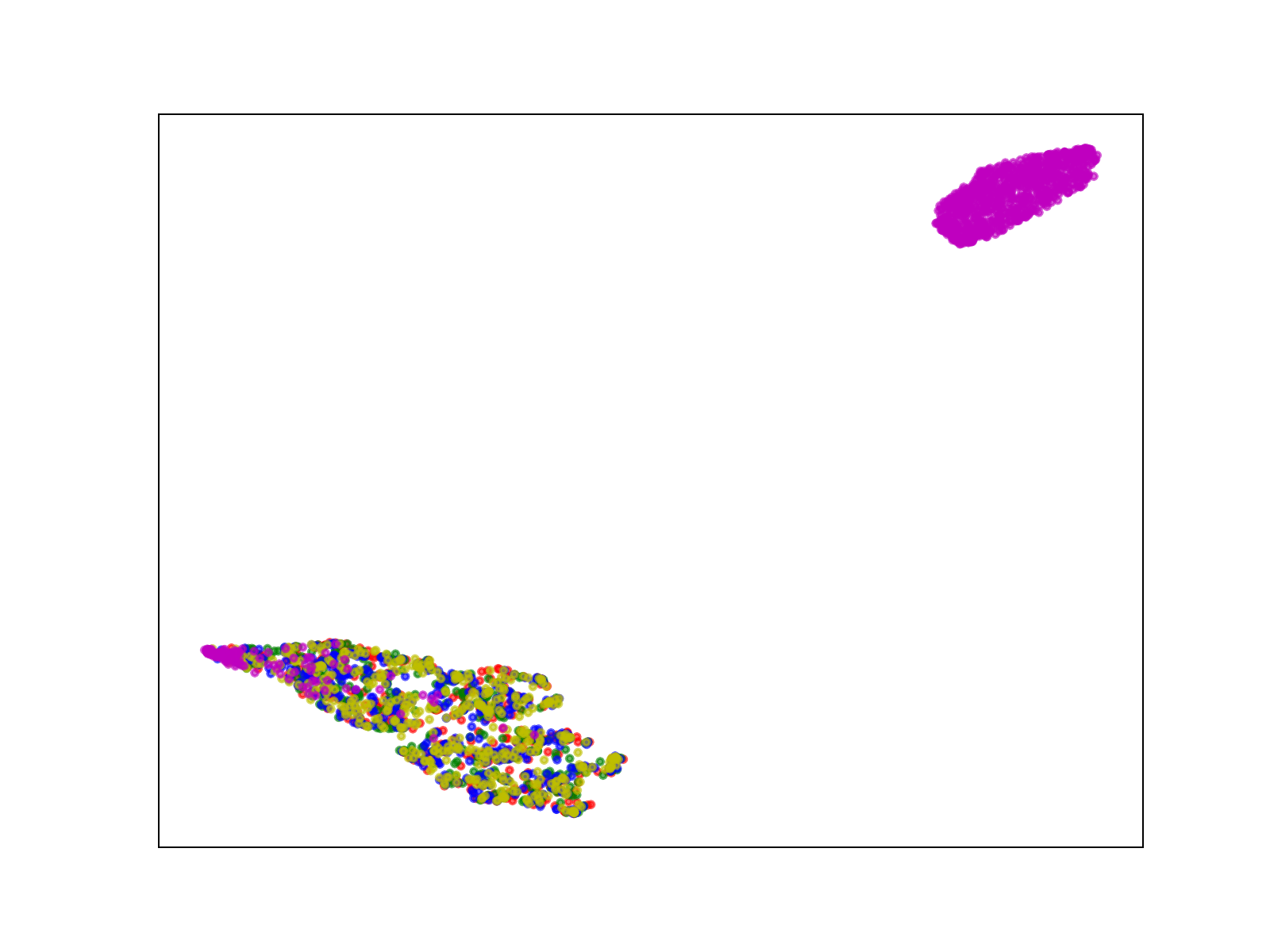}
        \caption{UMAP projection of swap data obtained from the toric code
        on a $20 \times 20$ lattice for all four topologically degenerate ground
        states and an excited state at excitation density $n_{ex} = 20\%$. 
        The projection was trained using only data from the ground state consisting
        of only homologically trivial loops and the excited state. Then we 
        subsequently apply the projection to the other three ground states.
        The purple dots are data from the excited state, the other colors are 
        from the ground states. 
        The overall accuracy is $98.04\%$.
        All of the groundstateable data cluster together and 
        more importantly, cluster separately from the excited state excepting the small
        fraction of errors.
        }
        \label{fig:tc_multigs}
    \end{figure}
    
        We now turn to gerneralizability. 
        Due to topological degeneracy, we have access to four groundstateable wave functions from the
        toric code. In fig.~\ref{fig:tc_multigs} we show the results of training the UMAP projection mapping
        for a $L = 20$ lattice using the groundstateable wave function containing only homologically trivial
        loops and the non-groundstateable wave function with $n_{ex} = 20\%$. We then transfer the projection map
        to swap data obtained from the other three groundstateable wave functions (those with an odd parity of 
        non-contractible loops around one or both cycles of the torus). 
        In fig.~\ref{fig:tc_multigs}, we can see that the data from the
        non-groundstateable wave function (purple dots) clusters separately from the groundstateable data (other colors),
        which all clusters together. 
        The accuracy of the collective projection is $98.04\%$, compared to $95.1\%$ from the initial data used
        to train the projection map. This makes sense because the only errors are non-groundstateable data being
        classified as groundstateable, so adding more groundstateable data reduces the error.
        This shows that the learned UMAP projection trained on one ground state generalizes to other ground states in the presence of topological degeneracy.  
        
    Another interesting feature of the clustering in this case is that
    misclassifications are always excited states being incorrectly
    classified as ground states.
    The distinction between the ground state and excited state is the presence of 
    spinons and the string operators connecting them. 
    To detect the excited nature of the wave function, a swap operator must swap a subsystem 
    in a way that cuts a string operator. 
    We therefore conjecture that misclassifications of 
    MC samples from excited states as ground states is due to VMC configurations in which the string
    operators connecting spinons are sufficiently short such that very few
    subsystems pick up the excited character of the wave function.
    
\section{Conclusion}
 
    In summary we introduced EntanCl, an unsupervised machine learning method to separate out the ground-stateable wave functions from the
    exponentially large Hilbert space of many-body wave functions with high computational efficiency.
    EntanCl consists of three steps: (1) preparation of input data, (2) projection of the data down to two-dimensional space using UMAP, (3) K-means clustering of the projected data.   
    The input data of our choice are
    matrix elements of an ensemble of
    swap operators collected as snapshots of individual uncorrelated variational Monte Carlo steps.
    By using the noisy snapshots as opposed to demanding convergence of the swap operator expectation value, EntanCl gains computational efficiency. We applied EntanCl to  a simple 
    one-dimensional band insulator model and from Kitaev's toric code to find accuracte clustering results. Moreover, we established that the learned UMAP projection is generalizable to an expansion of the data set. The clustering errors are found to occur asymmetrically: an excited state may get misplaced into the ground state cluster but not vice versa. Hence the cluster assignment into excited states will be a reliable way of ruling out groundstateability of the quantum many-body state. 
    As with any VMC sampling, the quality of the results can depend on the sampling basis due to the basis dependence in the spread of the noise. As we demonstrate in appendix B, as long as the spread of the noise remains comparable under a basis transformation, EntanCl will work independent of the basis choice.
    
    In the same vein of addressing wave functions, a more ambitious approach would be to attempt to reconstruct the Hamiltonian that takes a given wave function as its ground state. There has been recent progress in this direction with concrete proposals \cite{qi:q2019a,bairey:prl2019a,chertkov:pr2018a,garrison:pr2018a}. However, the Hamiltonian reconstruction is computationally costly as it requires precise measurements of many correlation functions. EntanCl can be a swift first pass that can weed out non-groundstateable many-body states without reference to Hamiltonians. 
    Furthermore, as a method that can efficiently sort the swap data associated with different quantum many-body states based on the their entanglement structure, we anticipate EntanCl to find applications beyond separating out ground-stateable wavefunctions. For instance, EntanCl will be ideal for studying quantum phase transitions involving change of entanglement structure due to spontaneous symmetry breaking or topological order \cite{metlitski:ap2015a}.

{{\bf Acknowledgements:} E-AK and MM are supported by the U.S. Department of Energy, Office of Basic Energy Sciences, Division of Materials Science and Engineering under Award DE-SC0018946 Grant. YZ is supported by the startup grant at Peking University. TS is supported by a US Department of Energy grant
DE- SC0008739, and in part by a Simons Investigator
award from the Simons Foundation.  TS is also supported by the Simons Collaboration on Ultra-Quantum Matter, which is a grant from the Simons Foundation (651440, ST).”The project was initiated at Kavli Institute of Theoretical Physics supported by the National Science Foundation under Grant No. NSF PHY-1748958.}

\bibliographystyle{apsrev4-1}
\bibliography{library}

\appendix
\section{\\ Overview of UMAP Procedure}
The purpose of the uniform manifold approximation and projection (UMAP) algorithm
is to create a low-dimensional projection of high-dimensional data such that the nearest
neighbors of a data point in high dimensions remain its nearest neighbors in the low
dimensional projection. How many nearest neighbors we try to keep is an input parameter
to the algorithm. This is useful for us because data that belong to distinct clusters in the
high dimensional space will not share nearest neighbors between clusters. Thus, in the
low-dimensional space, these data should still show up as distinct clusters. Here
we give an overview of how this algorithm works.
\begin{enumerate}
	\item Let $X = \{X_1,\dots,X_N\}$ denote our set of input data where
	each $X_i$ is an $n$-dimensional vector. Let
	$Y = \{Y_1,\dots,Y_N\}$ denote the output projected data points where
	$Y_i$ corresponds to the projection of $X_i$ and
	each $Y_i$ is a $d$-dimensional vector with $d \leq n$.  

	\item We would like the data to be uniformly distributed on the underlying manifold
	because then  the collection of local neighborhoods of our
	data points provide a good picture of the underlying manifold.
	UMAP \emph{forces} our data to be
	uniformly distributed by normalizing the distance from each point to the
	furthest neighbor we would like to consider. We are also going to assume that there
	are no isolated points on the underlying manifold, which we will enforce by
	fixing the distance to the \emph{nearest} neighbor. To do this,
	we define a local metric $d_i$ for each input data point $X_i$
	\begin{align*}
		d_i(X_j,X_k) = \begin{cases} \frac{1}{r_i}d_{\mathbb{R}^n}(X_j,X_k) - \rho_i &
		\text{if } i = j \text{ or } i = k \\
		\infty & \text{otherwise} \end{cases}
	\end{align*}
	where $d_{\mathbb{R}^n}$ is the Euclidean metric on $\mathbb{R}^n$, $\rho_i$ fixes the distance to the nearest neighbor to be zero, 
	and $r_i$ fixes the distance to the furthest neighbor we would
	like to consider. Note that we choose $r_i$'s so that for each $d_i$, the distance
	from $X_i$ to its furthest relevant neighbor is the same.
	For the projected output, we will define local metrics as well. The difference in 
	the projected space 
	is that we know what the underlying manifold is ($\mathbb{R}^d$) so we know what the
	true metric is. UMAP still enforces an assumption of local connectivity.
	Our local metrics for the encoded output $Y_i$'s are then
	\begin{align*}
		d_i(Y_j,Y_k) = \begin{cases} d_{\mathbb{R}^d}(Y_j,Y_k) - \rho_i &
		\text{if } i = j \text{ or } i = k \\
		\infty & \text{otherwise} \end{cases}
	\end{align*}

	\item Comparisons of distance between our different local metrics are meaningless,
	which seems to give us no way to assess the quality of a projection. To circumvent
	this UMAP considers a new represendation of the data:
	a \emph{neighborhood graph}. To build the graph, UMAP draws an edge between each
	data point and each of its neighbors up to the furthest one we would like to consider.
	The edges are weighted, where for an edge from $X_i$ to $X_j$, the
	weight of the edge is $\exp(-d_i(X_i,X_j))$. UMAP performs the same procedure
	for the projected data $Y$. Note that $d_i(X_i,X_j)$ is not neccesarily
	equal to $d_J(X_j,X_i)$. Thus, the edges drawn between $X_i$ and $X_j$ 
	by $d_i$ and $d_j$ may not have the same weight. 

	\item Next UMAP combines edges so that there is at most one edge between
	any two points. The edges are combined pairwise
	where for a pair of edges with weights $\alpha,\beta$, UMAP forms a combined edge with
	weight $f(\alpha,\beta) = \alpha+\beta - \alpha \cdot \beta$. This process occurs
	for both the input data $X$ and the projected data $Y$. The function $f$ is not
	the unique way to combine edge weights, but is a choice made by UMAP.

	\item Now we have a neighborhood graph for $X$ and $Y$ with an unambiguous
	definition of \emph{the} edge between two points. Because the neighborhood graphs
	for $X$ and $Y$ have the same number of vertices and each vertex is the same
	degree, we can define an isomorphism between them. We do this by associating 
	projected points with data points being careful to ensure that if there is an
	edge between $X_i$ and $X_j$, the points $Y_i$ and $Y_j$ that we associate with them
	are also connected by an edge. Thus we can speak unambiguously about
	a single edge set $E$.
	To measure the "similarity"
	of the two neighborhood graphs, we will use the cross entropy
	\begin{align*}
	C(E;\mu_\cup,\nu_\cup) \equiv \sum\limits_{e \in E} \mu_\cup(e)
	\log\left(\frac{\mu_\cup(e)}{\nu_\cup(e)}
	\right) + \\(1- \mu_\cup(e))\log\left(\frac{1-\mu_\cup(e)}{1-\nu_\cup(e)}\right)
	\end{align*}
	where $E$ is the set of edges, $\mu_\cup(e)$ is the combined weight
	(as in step 4)
	of an edge in $Y$, and $\nu_\cup(e)$ is the combined weight of an edge
	in $X$.
	We can minimize the cross entropy using stochastic gradient descent. For each
	step of the optimization we move the positions of the encoded points, changing
	the distance, and therefore the edge weights, between them. 
\end{enumerate}

\section{\\ Example of Basis Dependence}
\begin{figure}
    \centering
    \includegraphics[width=\linewidth]{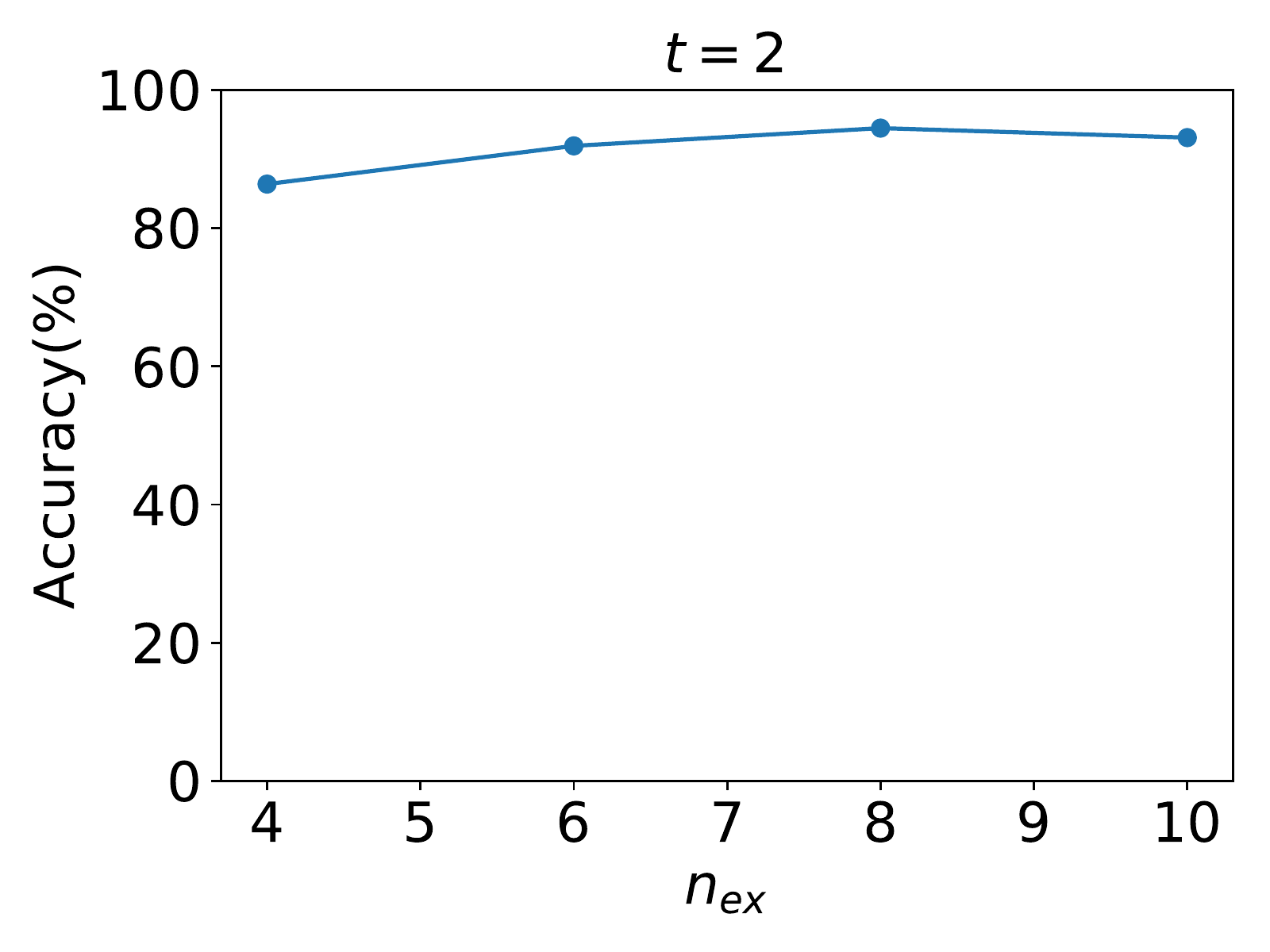}
    \caption{
    Here we show the clustering accuracy for swap data obtained from the
    ground state wavefunction of $\mathcal{H}_k'$ (c.f. eq.~\ref{eqn:basis_model}) and 
    non-groundstateable wave functions with normalized energy gap $t = 2$ and varying
    excitation density $n_{ex}$. Although the accuracy at similar $n_{ex}$ is lower for the
    model in this basis than the original (c.f. fig.~\ref{fig:band_insulator}(b)), the accuracy
    is still high (peaking over 90\%) and stays above 80\% even at low $n_{ex}$ values.
    }
    \label{fig:basis_dependence}
\end{figure}

A basis transformation can affect the spread in the VMC data obtained during step one of EntanCl
by changing the relative magnitudes of the coefficients $C_{\alpha\beta}$ in the wave function 
(c.f. eq.~\ref{eqn:matelt}).
This change in the spread of the data can affect the accuracy of the resultant clustering if the neighborhoods of MC samples from groundstateable  wave  functions  intersect  those  of  non-groundstateable wave functions in the high dimensional space.
Here we discuss an example of the basis dependence of our results by re-examining the
band insulator model of section III under a basis transformation.
The $k$-space Hamiltonian for the original band insulator model is given by
\begin{align}
    \mathcal{H}_k = [ t_1 + t_2 \cos(k) ] \sigma^x - t_2 \sin(k) \sigma^y
\end{align}
where the $\sigma^i$'s are Pauli matrices. We now consider a new model that differs
from the original by an $SU(2)$ unitary transformation with Hamiltonian
    \begin{eqnarray}
        \mathcal{H}' &=& \sum\limits_i t_1 (a_i^\dagger a_i - b_i^\dagger b_i)  \\& &+ \frac{t_2}{2}(a_{i+1}^\dagger a_i-b_{i+1}^\dagger b_i + b_{i+1}^\dagger a_i - b_{i-1}^\dagger a_i + \text{h.c.}) \nonumber
    \end{eqnarray}

\begin{align}
    \label{eqn:basis_model}
    \mathcal{H}_k' = [t_1+t_2\cos(k)]\sigma^z + t_2 \sin(k)\sigma^y.
\end{align}
This new model $\mathcal{H}_k'$ describes the same physics as $\mathcal{H}_k$, but
differs by a basis transformation. We show the clustering accuracy results of scaling
the excitation density $n_{ex}$ at fixed normalized gap $t = 2$ in
fig.~\ref{fig:basis_dependence}. We can see that, as was the case in 
fig.~\ref{fig:band_insulator}, the accuracy is high and remains high even at low 
$n_{ex}$ values. However, the accuracy in this basis is not as high as in the original
basis at the same $n_{ex}$ values. This illustrates that noise in the VMC data does
indeed carry a basis dependence, but that sampling data in a new basis does not necessarily
destroy the separability of the swap data from groundstateable and non-groundstateable wave
functions. 

\end{document}